\documentclass[conference]{IEEEtran}
\IEEEoverridecommandlockouts
\usepackage{cite}
\usepackage{amsmath,amssymb,amsfonts}
\usepackage{algorithmic}
\usepackage{graphicx}
\usepackage{url}
\usepackage[T1]{fontenc}
\usepackage{textcomp}
\usepackage{xcolor}
\def\BibTeX{{\rm B\kern-.05em{\sc i\kern-.025em b}\kern-.08em
    T\kern-.1667em\lower.7ex\hbox{E}\kern-.125emX}}

\usepackage{moresize}
\usepackage{listings}

\definecolor{darkgreen}{rgb}{0.0,0.4,0.0} 
\definecolor{darkred}{rgb}{0.6,0.1,0.1}
\definecolor{lightgray}{gray}{.98}
\definecolor{medgray}{gray}{.70}
\definecolor{darkgray}{gray}{.40}
\definecolor{lightviolet}{rgb}{0.7,0,0.7} 
\definecolor{lightlightviolet}{rgb}{1.0,0.7,1.0} 
\definecolor{darkviolet}{rgb}{0.5,0.1,0.5}
\definecolor{darkredviolet}{rgb}{0.6,0.1,0.4}
\definecolor{limegreen}{rgb}{0.2,0.7,0.2}
\definecolor{navyblue}{RGB}{0,0,128}
\definecolor{aquamarine}{RGB}{102,205,170}

\definecolor{strictRED}{RGB}{184,0,0}
\definecolor{specificationTURQUOISE}{RGB}{0,128,153}
\definecolor{assumptionGREEN}{RGB}{0,128,0}
\definecolor{interruptBLUE}{RGB}{0,0,128}
\definecolor{committedORCHID}{RGB}{54,22,89}
\definecolor{urgentORCHID}{RGB}{74,28,109}
\definecolor{requestedORCHID}{RGB}{104,34,139}
\definecolor{eventuallyORCHID}{RGB}{154,50,205}

\lstdefinelanguage{SMLX}
{
	basicstyle=\ssmall\ttfamily, %
	frame=single, 
	framextopmargin=0pt,
	framexbottommargin=0pt,
	framexleftmargin=0pt,
	xleftmargin=16pt,
	xrightmargin=3pt,
	morekeywords=[1]{system, domain, scenario, bind, to, 
		message, non, spontaneous, events, specification, 
		alternative, if, collaboration, role, with, dynamic, 
		bindings, or, and, null, define, as, 
		constraints, import, static, parameter, ranges, var, EInt, 
		controllable},
	morekeywords=[2]{strict},
	morekeywords=[3]{forbidden, violation},
	morekeywords=[4]{interrupt},
	morekeywords=[5]{guarantee},
	morekeywords=[6]{assumption}, 
	morekeywords=[7]{committed}, 
	morekeywords=[8]{urgent},
	morekeywords=[9]{requested},
	morekeywords=[10]{eventually},
	keywordstyle=[1]\color{darkviolet}\textbf,
	keywordstyle=[2]\color{strictRED}\textit,
	keywordstyle=[3]\color{strictRED}\textit,
	keywordstyle=[4]\color{interruptBLUE}\textit,
	keywordstyle=[5]\color{specificationTURQUOISE}\textbf,
	keywordstyle=[6]\color{assumptionGREEN}\textbf,
	keywordstyle=[7]\color{committedORCHID}\textit,
	keywordstyle=[8]\color{urgentORCHID}\textit,
	keywordstyle=[9]\color{requestedORCHID}\textit,
	keywordstyle=[10]\color{eventuallyORCHID}\textit,
	sensitive=false,
	morecomment=[l][\color{darkgreen}\textit]{//},
	morecomment=[s][\color{darkgreen}\textit]{/*}{*/}, 
	morestring=[b][\color{blue}]",
	tabsize=1,
	moredelim = [s][\color{specificationTURQUOISE}\textbf]{guarantee}{scenario},
	moredelim = [s][\color{assumptionGREEN}\textbf]{assumption}{scenario},
	backgroundcolor=\color{lightgray}
}

\lstdefinestyle{SMLXStyle} {language=SMLX}

\lstdefinelanguage{SMLConfig}
{
	basicstyle=\ssmall\ttfamily,
	frame=single, 
	framextopmargin=0pt,
	framexbottommargin=0pt,
	framexleftmargin=0pt,
	xleftmargin=16pt,
	xrightmargin=3pt,
	morekeywords=[1]{symbolic, import, configure, specification, use, 
		instancemodel, symbolic, parameters, attributes, symbolic, state, matching, 
		off, under, approximation, on, rolebindings, collaboration, object, plays, 
		role, role1},
	keywordstyle=[1]\color{darkviolet}\textbf,
	sensitive=false,
	morecomment=[l][\color{darkgreen}\textit]{//},
	morecomment=[s][\color{darkgreen}\textit]{/*}{*/}, 
	morestring=[b][\color{navyblue}\textit]",
	stringstyle=\color{navyblue},
	tabsize=1,
	backgroundcolor=\color{lightgray}
}

\lstdefinestyle{SMLConfigStyle} {language=SMLConfig}
\lstdefinelanguage{Java}
{
	basicstyle=\ssmall\ttfamily,
	frame=single, 
	framextopmargin=0pt,
	framexbottommargin=0pt,
	framexleftmargin=0pt,
	xleftmargin=16pt,
	xrightmargin=3pt,
	morekeywords=[1]{public, private, class, extends, protected, void,
		new, throws, null, if, else},
	morekeywords=[2]{STRICT},
	morekeywords=[3]{@Override},
	morekeywords=[4]{car, oc, cp}, 
	keywordstyle=[1]\color{darkviolet}\textbf,
	keywordstyle=[2]\color{javablue}\textbf,
	keywordstyle=[3]\color{darkgray},
	keywordstyle=[4]\color{navyblue},
	sensitive=false,
	morecomment=[l][\color{javagreen}\textit]{//},
	morecomment=[s][\color{javagreen}\textit]{/*}{*/}, 
	morestring=[b][\color{javablue}\textit]",
	stringstyle=\color{navyblue},
	tabsize=1,
	backgroundcolor=\color{lightgray}
}

\definecolor{javared}{rgb}{0.6,0,0} 
\definecolor{javablue}{rgb}{0,0,0.9} 
\definecolor{javagreen}{rgb}{0.25,0.5,0.35} 
\definecolor{javapurple}{rgb}{0.5,0,0.35} 
\definecolor{javadocblue}{rgb}{0.25,0.35,0.75} 

\lstdefinestyle{JavaStyle} {language=Java}

\lstdefinelanguage{Kotlin}{
	basicstyle=\ssmall\ttfamily,
	frame=single, 
	framextopmargin=0pt,
	framexbottommargin=0pt,
	framexleftmargin=0pt,
	xleftmargin=16pt,
	xrightmargin=3pt,
	comment=[l]{//},
	commentstyle={\color{darkgray}\ttfamily},
	emph={delegate, filter, first, firstOrNull, forEach, lazy, map, mapNotNull, println, return@, event, sends, request},
	emphstyle={\color{darkviolet}},
	identifierstyle=\color{black},
	numberstyle=\color{darkgreen},
	keywords=[1]{ abstract, actual, as, as?, break, by, companion, continue, data, do, dynamic, else, enum, expect, false, final, for, get, if, import, in, interface, internal, is, null, object, override, package, private, public, return, set, super, suspend, this, throw, true, try, typealias, val, var, vararg, when, where, while},
	keywordstyle=[1]{\color{javablue}\bfseries},
	keywords=[2]{@Deprecated, @JvmField, @JvmName, @JvmOverloads, @JvmStatic, @JvmSynthetic, @Test, Array, Byte, Double, Float, Int, Integer, Iterable, Long, Short, String, scenario, class, fun},
	keywordstyle=[2]{\color{javablue}},	
	keywords=[3]{interruptingEvents, forbiddenEvents, it}, 
	keywordstyle=[3]{\color{darkviolet}\bfseries},
	keywords=[4]{scenario, cycleScenario, runTest, Given, When, Then, And, But}, %
	keywordstyle=[4]{\textit},
	keywords=[5]{coolantTemp, deratingFactor}, 
	keywordstyle=[5]{\color{darkviolet}\bfseries\underbar},
	keywords=[6]{currentTemp}, 
	keywordstyle=[6]{\underbar},
	morecomment=[s]{/*}{*/},
	morecomment=[s][\color{black}]{`}{`},
	morestring=[b]",
	morestring=[s]{"""*}{*"""},
	sensitive=true,
	stringstyle={\color{javagreen}\ttfamily},
}

\lstdefinestyle{KotlinStyle} {language=Kotlin}

\newcommand{\lstinlineKotlin}[1]{\lstinline[language=Kotlin,basicstyle=\small\ttfamily]{#1}}

\lstdefinelanguage{Gherkin}{
	basicstyle=\ssmall\ttfamily,
	frame=single, 
	framextopmargin=0pt,
	framexbottommargin=0pt,
	framexleftmargin=0pt,
	xleftmargin=16pt,
	xrightmargin=3pt,
	comment=[l]{//},
	commentstyle={\color{darkgray}\ttfamily},
	emph={@software, @charging },
	emphstyle={\color{limegreen}},
	identifierstyle=\color{black},
	keywords={Given, When, Then, And},
	keywordstyle={\color{violet}\ttfamily},
	morecomment=[s]{/*}{*/},
	morestring=[b]",
	morestring=[s]{"""*}{*"""},
	ndkeywords={Scenario, Example, Feature},
	ndkeywordstyle={\color{darkviolet}\bfseries},
	sensitive=true,
	stringstyle={\color{javagreen}\ttfamily},
}

\lstdefinestyle{GherkinStyle} {language=Gherkin}

\makeatletter
\lstdefinestyle{mystyle}{
	basicstyle=%
	\ttfamily
	\lst@ifdisplaystyle\footnotesize\fi
}
\makeatother

\begin{document}

\title{Selecting Features for the Next Release\\ in a System of Systems Context}

\makeatletter
\newcommand{\linebreakand}{%
  \end{@IEEEauthorhalign}
  \hfill\mbox{}\par
  \mbox{}\hfill\begin{@IEEEauthorhalign}
}
\makeatother

\author{\IEEEauthorblockN{Carsten Wiecher}
\IEEEauthorblockA{
\textit{{Dortmund University of Applied }}\\
\textit{{Sciences and Arts}}\\
\textit{44139 Dortmund, Germany}\\
carsten.wiecher@fh-dortmund.de}
\and
\IEEEauthorblockN{Carsten Wolff}
\IEEEauthorblockA{
\textit{{Dortmund University of Applied }}\\
\textit{{Sciences and Arts}}\\
\textit{44139 Dortmund, Germany}\\
carsten.wolff@fh-dortmund.de}
 \linebreakand
 \IEEEauthorblockN{Harald Anacker}
\IEEEauthorblockA{\textit{Fraunhofer IEM} \\
\textit{33102 Paderborn, Germany}\\
harald.anacker@iem.fraunhofer.de}
\and
\IEEEauthorblockN{Roman Dumitrescu}
\IEEEauthorblockA{\textit{Fraunhofer IEM} \\
\textit{33102 Paderborn, Germany}\\
roman.dumitrescu@iem.fraunhofer.de}}


\maketitle

\vspace{-2cm}
\begin{abstract}
Smart Cities are developing in parallel with the global trend towards urbanization. The ultimate goal of Smart City projects is to deliver a positive impact for the citizens and the socio-economic and ecological environment. This involves the challenge to derive concrete requirements for (technical) projects from overarching concepts like Quality of Life (QoL) and Subjective Well-Being (SWB). Linking long-term, impact oriented goals with project outputs and outcomes is a complex problem. Decision making on requirements and resulting features of single Smart City projects (or systems) is even more complex since cities are not like monolithic, hierarchical and well-structured systems. Nevertheless, systems engineering provides concepts which support decision making in such situations. Complex socio-technical systems such as smart cities can be characterized as systems of systems (SoS). A SoS is composed of independently developed systems that nevertheless provide a higher-level integrated functionality. To add new functionality to a SoS, either existing systems must be extended or new systems must be developed and integrated. In both cases, the extension of functionality is usually done in small increments and structured via software releases. However, the decision which features to include in the next release is complex and difficult to manage when done manually. To address this, we make use of the multi-objective next release problem (MONRP) to search for an optimal set of features for a software release in a SoS context. In order to refine the search 
in an early planning phase, we propose a technique to model and  validate the features using the scenario modeling language for Kotlin (SMLK). This is demonstrated with a proof-of-concept implementation.
\end{abstract}


\section{Introduction}
Decision making, project selection and goal-setting for projects are a very relevant long time issue in politics and city planning. E.g., in Germany, there is a long history of attempts, methods and tools in order to manage the development of cities and the political decision making \cite{Zabel2020}. Several programs with top-down planning and decision making where implemented. It is an ongoing debate if they met their goal and if they had a clear goal. Cities are highly complex, highly connected and organic, self-organizing systems. The digital transformation towards Smart Cities is adding to this complexity. Decisions on which steps to take next, which projects to do, which goals to set for the projects and which features to implement are difficult. This makes the requirements engineering and project management for engineering projects in the Smart City context very difficult, too. This contribution intends to deliver a decision support approach which is based on systems engineering methodology. It is believed that the approach supports the effort to link impact-oriented indicator systems for Smart Cities like Quality of Life (QoL) or Subjective Well-Being (SWB) to the requirements for concrete engineering projects. Complex systems like smart cities can be characterized as system of systems (SoS) \cite{Lane2013WhatIA}. These SoS are composed of individual constituent systems (CSs) that are independently developed and operated but nevertheless can be integrated to provide a higher level functionality \cite{Maier1996}. One use case in a Smart City context is smart charging of electric vehicles (EV). To realize a smart charging use case, different systems from different domains (energy conversion, energy transfer from/to EV, EV user premises etc.) must be considered and integrated~\cite{Kirpes2019}. 

In contrast to the development of complex monolithic systems, new functionality in a SoS context is usually created by selecting, adapting and integrating systems \cite{Dahmann2008}. Since these systems are independently operated and managed, can have different life cycles and evolve over time, the decision which functionality shall be added is a complex task. It is necessary to make trade-off decisions based on expected costs and values for the realization of new functionalities. The estimation of the value of a new feature in the context of Smart City projects is particularly challenging as we describe in section II. Therefore, the indicators of cost and value of new functionalities form the interface to the decision-making processes of the Smart City.   

In this context, we address the problem that in a SoS we do not have a fixed set of requirements that we can use to decompose the system requirements into subsystem requirements \cite{Ncube2011} \cite{Ncube2018} in order to estimate the cost and value for realizing these requirements.

In this paper, we propose an iterative and tool-supported process that helps the release engineer to identify a set of features that can be considered for a next release to add new functionality to a SoS. 
We use a software release to group a set of features, where features are used to summarize a consistent set of requirements from one or more stakeholders. 

As introduced in previous work \cite{Wiecher2019, Wiecher2020, Wiecher2020a}, we show how to specify stakeholder expectations in a comprehensive and structured form using feature files and usage scenarios. Based on this feature specification, we use a test-driven specification approach \cite{Wiecher2019} to iteratively model the intended SoS behavior in form of a scenario specification \cite{Wiecher2021}. By creating, executing and testing the scenario specification, we are able to refine the initial feature specification. 

Subsequently we use this feature specification along with the scenario specification to manually estimate the costs it can take to realize the feature and the value it can create. 
This estimation is used as input to a multi objective next release problem (MONRP) \cite{Zhang2007a} that we apply to automatically search for an optimal set of features. The resulting feature sets of possible release candidates can be used by the release engineer to select the next release.

We show a proof-of-concept implementation, where features are specified using the Gherkin syntax \cite{SmartBearSoftware} within the Cucumber tooling \cite{cucumber}. The SoS behavior is modeled using the scenario modeling language for Kotlin (SMLK)\cite{smlk}, and the MONRP is implemented using the MOEA framework~\cite{Hadka}. 
As a result, we see that the iterative process of 1) structured feature specification, 2) test-driven and scenario-based requirements modeling, and 3) a tool-supported search for release candidates, can be supportive to identify a next release in a SoS context. 

\section{Background}
\subsection{Smart City projects}
Citizens expect from Smart City projects a positive impact. Nevertheless, it is difficult to define this positive impact clearly and it is even more difficult to link it to concrete events or elements within a Smart City. E.g., there are attempts to link concepts like Quality of Urban Life (QOUL) and Subjective-Well Being (SWB) to concrete features of a city~\cite{Ala-Mantila2018}. This involves two underlying problems: one is the definition and measurement of indicator sets describing QOUL or SWB. The second issue is the linking of long-term impact goals like QOUL or SWB to concrete project outputs which can be described by user requirements. Systems engineering methodology requires such well-defined user requirements in order to develop features of technical systems. The systematic linking of inputs, outputs, outcomes and impact of projects is a relevant research topic in project management \cite{Turner2012}. The cost of a new functionality in the sense of this study is derived from the inputs required by the respective project. The value is addressing the impact which is difficult to determine at the time of decision-making. This can be addressed by a result-oriented logic defining potential cause and effect chains between the different levels of results. Nevertheless, such logic remains incomplete and insufficient in complex real-world scenarios like Smart Cities due to the high level of uncertainty. This leads to a more incremental process of defining and selecting projects which add systems to the Smart City step-by-step. Decision making has to focus on selecting projects and Smart City features and functionalities which are implemented and released to the public. The management and decision-making for the release of such new features can be supported by systems engineering methodology. Doing "good" decisions can be supported by optimization methods from operations research. The following sections describe how features can be selected for inclusion into the release of a new system and how these systems can be integrated into the overall Smart City system.   

\subsection{System of Systems}
\label{sect:sos}
SoS are distinguished from complex monolithic systems by the operational, managerial and evolutionary independence of its CSs \cite{Maier1996}. Accordingly, each CS can perform a meaningful task, even if it's not integrated into the SoS. Since the CSs are self-administered and individually managed, this can lead to conflicting development goals between the single CSs, or between CSs and the SoS. Also,  objectives and functionality of an SoS can change constantly, as they can be added, modified or removed based on experience. Therefore an SoS never appears to be fully completed  \cite{Nielsen2015}.

Due to these SoS characteristics and in contrast to the development of complex monolithic systems, we do not have a fixed set of requirements that can be decomposed for an linear and top-down system design \cite{Ncube2011}\cite{Ncube2018} and validation~\cite{Honour2013}. However, requirements decomposition techniques and the alignment of stakeholder expectations via validation are mandatory to estimate the value and the costs of requirements and to decide which requirements should be realized with the next release. 
\begin{figure}[h]
    \centering
    \includegraphics[width=1.0\linewidth]{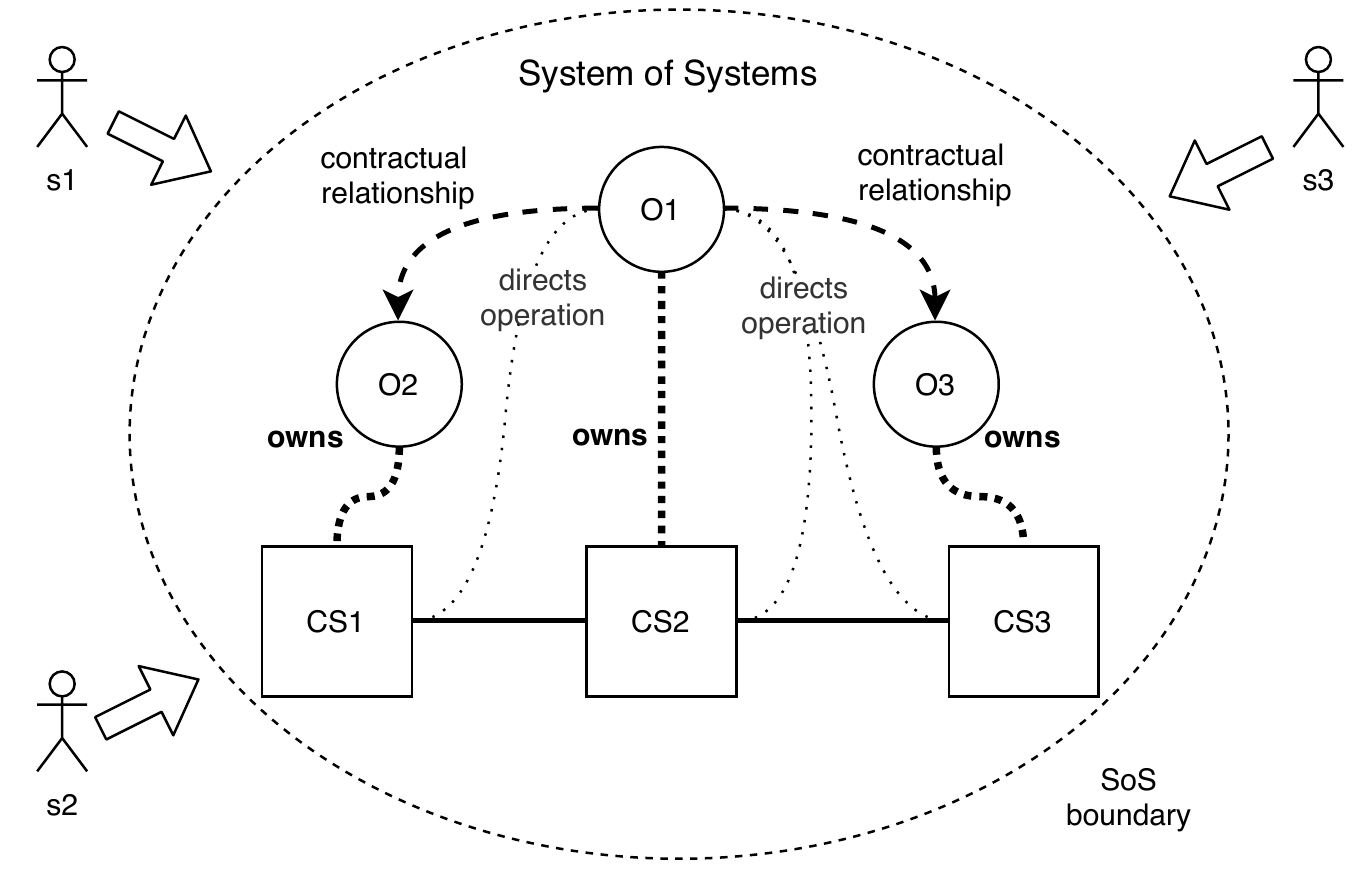}
    \caption{Acknowledged System of Systems, based on \cite{Dahmann2008} \cite{Ncube2018}}
    \label{fig:systemOfSystems}
\end{figure}

To support SoS requirements engineering, Ncube and Lim argue that it's necessary to define a SoS type in an early development phase \cite{Ncube2018}. This is fundamental to understand the dependencies and responsibilities between the SoS and its CSs, and also between the individual CSs owner and other stakeholder during SoS development \cite{Ncube2018} \cite{Ncube2011}. 
Therefor, in this paper we make use of an acknowledged SoS \cite{Maier1996} \cite{Dahmann2008} \cite{Nielsen2015} as outlined in Fig.~\ref{fig:systemOfSystems}, which is suitable to describe systems like Smart Cities \cite{Ncube2018}. Here, we assume that we have three CSs (CS1, CS2, CS3) with each CS having one system owner (O1, O2, O3),
who is responsible for the development and operation of the respective CS. 
We also see external stakeholders (s1, s2, s3), that have an interest in the SoS development (e.g. SoS user).
In this acknowledged SoS, O1 directs the operation and the choice of CSs. For this, O1 can define requirements that must be fulfilled by the CSs. However, the CSs retain their independent ownership, funding, objectives, and development approaches. Hence, the realization of a SoS functionality is directed by a central authority but requires the collaboration on CS level \cite{Ncube2018}. 

\subsection{Multi-Objective Next Release Problem}
According to \cite{Franch2016}, a \emph{release} is a collection of new and/or changed features that form a new version of an evolutionary product/system. Thereby, a \emph{feature} serves to summarize a consistent bundle of requirements from one or more stakeholders~\cite{Chen2005}. 

If we characterize a Smart City as a SoS, the local government is an example of a central authority (O1 in Fig.~\ref{fig:systemOfSystems}) that defines the collection of features that should be included into the next release to form a new SoS. To support the central authority in deciding which features should be included into the next release, the application of the MONRP formulated in \cite{Zhang2007a} is suitable. 

As outlined in Fig.~\,\ref{fig:systemOfSystems} we have multiple stakeholders requesting new functionality from the SoS. On an SoS level, O1 has to select a set of features for the next release, where each feature is realized by a CS or by a combination of CSs. To realize a feature, each system owner has to spend an amount of resources resulting in a certain amount of costs. At the same time, when providing a specific feature, this provides a specific value. From the perspective of O1, the goal is to minimize the costs and maximize the value. As a result we have two conflicting objectives. According to \cite{Zhang2007a} and \cite{Papadimitriou1998}, finding the optimal set of features is a $\mathcal{NP}$-hard problem and cannot be resolved by exact optimization. For this reason, we use a metaheuristic search technique \cite{Zhang2007a} to approximate an optimal set of features.

\subsection{Scenario-based Requirements Modeling}
\label{sect:smlk}
In this paper, we argue that early requirements modeling and testing can support cost and value estimation of features to be implemented. Requirements modeling with scenarios and use cases gives stakeholders a better tool for decision-making since the scenarios and use cases can be formulated in natural language and they can be linked to real-life situations of the stakeholders. Therefore, on the one hand, requirements engineering based on scenarios and use cases offers a "common language" to involve non-technical stakeholders into the process. On the other hand, scenario and use case descriptions can be transformed into formal descriptions for requirements engineering by using model-based approaches. For this requirements modeling we use the scenario modeling language for Kotlin (SMLK), which is based on the concepts described in \cite{Harel2012, Damm2001, Greenyer2017}.  

In SMLK, functional requirements are modeled with behavioral threads which we call scenarios. These scenarios are loosely coupled via shared events. Within a scenario, we can request events that shall be executed. If an event is selected for execution, it can trigger the execution of other scenarios that in turn can request additional events.  
By iteratively adding scenarios to a scenario specification, we obtain an increasingly complete specification of the intended system behavior over time. During execution of the scenario specification as a scenario program, the scenarios are interwoven to produce a coherent system behavior that meets the requirements of all scenarios

An example scenario specification as part of a simplified smart charging use case is shown in Listing~\,\ref{list:bspSMLK}. 
\begin{lstlisting}[caption=SMLK scenario specification,
	label=list:bspSMLK,
	style=KotlinStyle
	]
class EVU{ // Electric Vehicle User
    fun energyPriceInformation() = event(){}
}
class App{ // Smartphone App
    fun enterChargingPreferences() = event(){}
    fun calculateChargingPlan() = event(){}
}
class EV{ // Electric Vehicle
    fun chargingPlan() = event(){}
    fun executeChargingPlan() = event(){}
}

scenario(EVU sends App.enterChargingPreferences()) {
    request(App.calculateChargingPlan())
    request(App sends EV.chargingPlan())   
    request(EV.executeChargingPlan())
}
\end{lstlisting}
In this specification we modeled two CSs (smartphone app and electric vehicle) and one external stakeholder (electric vehicle user). The scenario in line 13 is triggered when a electric vehicle user enters charging preferences to the app, i.e. the event \lstinlineKotlin{<EVU sends App.enterChargingPreferences()>} is executed. In this case, the body of the scenario is executed and further events are requested.  

Following the specification method we proposed in \cite{Wiecher2021}, we can intuitively model the SoS behavior on different levels of abstraction. Therefor, in a first step, we define the required CSs and the messages that these systems exchange (see Listing~\,\ref{list:bspSMLK}). In a second step, we refine the internal CS behavior by separate scenario specifications \cite{Wiecher2021}. 
 
\section{Feature Selection for the next Release}
We propose an iterative process as shown in Fig.~\ref{fig:steps}. 
The intent of this process is to support the definition of cost and value information that will serve as input to the automated search for an optimal set of features for the next release.
\begin{figure}[h]
    \centering
    \includegraphics[width=1.0\linewidth]{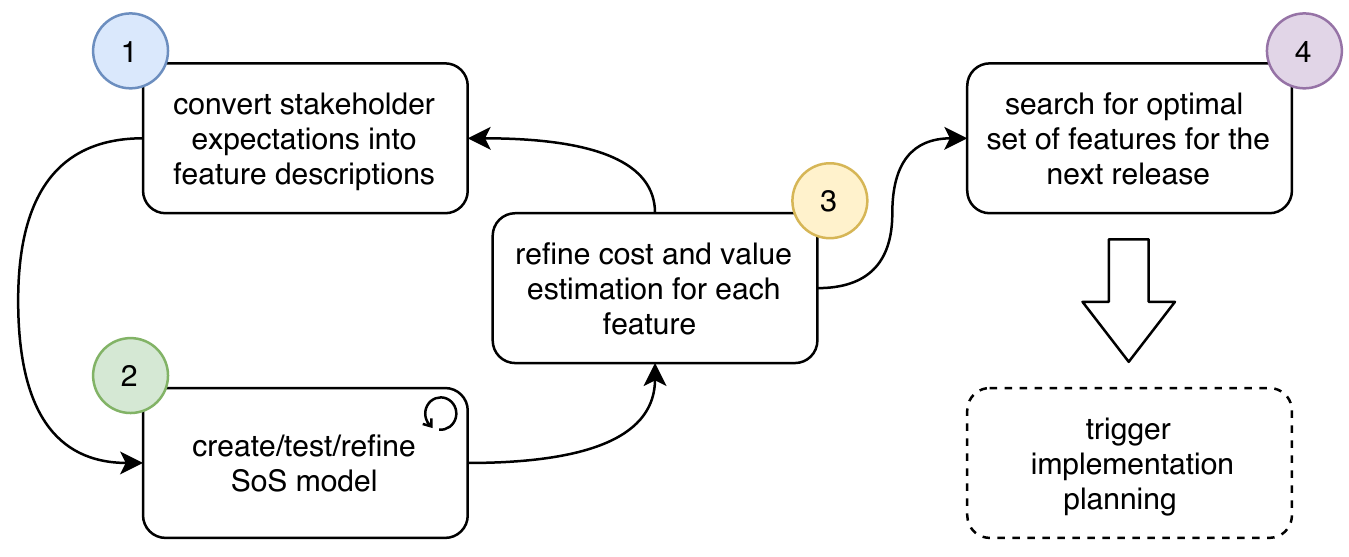}
    \caption{Steps for the tool supported selection of features for the next release.}
    \label{fig:steps}
\end{figure}
We start with a specification of features (1) which are manually derived from stakeholder expectations. Subsequently we model and execute the expected system behavior in short iterations (2) in order to 
elaborate a formal and scenario based specification of technical requirements. In (3) we iteratively refine the input data for the automated search which is executed in (4). 
As a result of this search, we get a set of features which can be used as a starting point for the implementation planning.

As outlined in Fig.~\,\ref{fig:valueAndCosts}, the process aims to link the business-level definitions of features with the specification of technical requirements. 
By utilizing usage scenarios, the artifacts in both views are connected with each other.
\begin{figure}[h]
    \centering
    \includegraphics[width=0.7\linewidth]{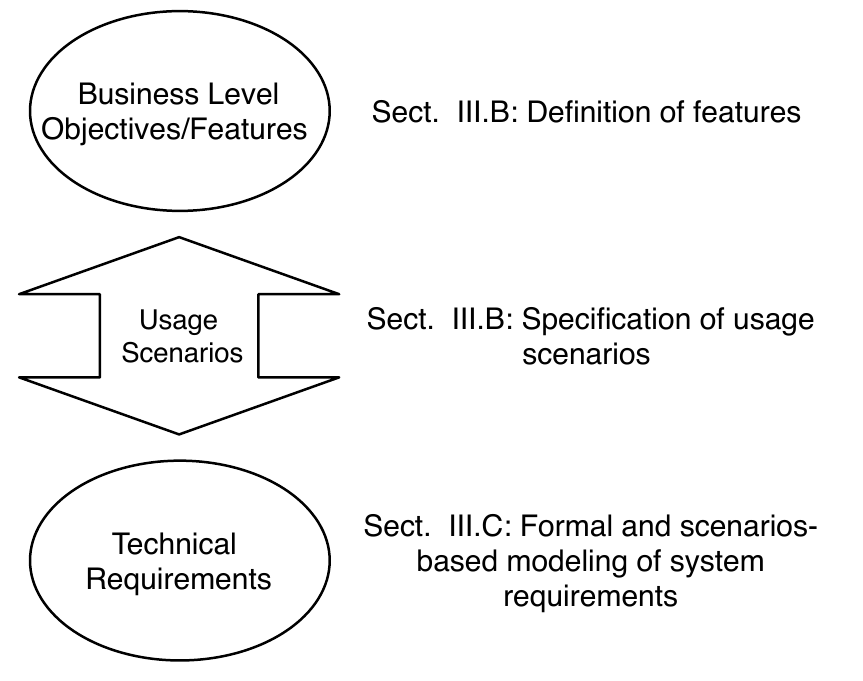}
    \caption{Definition of features and formal, scenario-based technical requirements, that are linked via usage scenarios.}
    \label{fig:valueAndCosts}
\end{figure}
In this way, features and technical requirements can be executed and refined iteratively. As a result, we obtain a feature specification that supports value estimation and a scenario-based  specification of technical requirements that supports estimation of the cost of implementing a particular feature.  

\subsection{Objective Functions}
\label{sect:selectFeaturesForTheNextRelease}
To perform an automated search for an optimal set of features for the next release, we use the MONRP as proposed in \cite{Zhang2007a}.

We assume that we have a set of stakeholders,
\begin{equation*}
S = \{s_1,...,s_m\}    
\end{equation*}
that have an legitimated interest in the realization of a new SoS functionality. Based on the desired SoS functionality, we derive a set of features,    
\begin{equation*}
F = \{f_1,...,f_n\}    
\end{equation*}
and for each feature $f_i(1\leq i \leq n)$ we define the estimated costs 
\begin{equation*}
Cost = \{c_1,...,c_n\}    
\end{equation*}
for the realization of that feature.

According to Fig.~\ref{fig:systemOfSystems}, we see the stakeholder outside of the SoS boundary. Within the SoS, the system owner O1 directs the composition of the individual CSs. From the perspective of this system owner, the importance of the individual stakeholder might vary. Consequently, as proposed in \cite{Zhang2007a}, the relative weight for each stakeholder $s_j = (1 \leq j \leq m )$ is denoted by: 
\begin{equation*}
Weight = \{w_1,...,w_m\}    
\end{equation*}
with $w_j \in [ \,0,1] \,$ and $\sum_{j=1}^{m} w_j = 1$. 

In addition, for each stakeholder $s_j(1 \leq j \leq m)$, we assign a value to a feature $f_i(1 \leq i \leq n)$ indicated with $value(f_i, s_j )$ where $value(f_i, s_j ) > 0$ if the stakeholder $j$ is interested in feature $i$ and $0$ otherwise.

The resulting importance of a feature is calculated with:

\begin{equation*}
score_i  = \sum_{j=1}^{m} w_j \cdot value(f_i, s_j)   
\end{equation*}

Based on this, we can define the two objective functions for the metaheuristic search. First, we want to maximize the overall value:  
\begin{equation*}
    Maximize \,\,\,\,\, f_1 ( \, \vec{x})\, = \sum_{i=1}^{n} score_i \cdot x_i  
\end{equation*}

and second, we want to minimize the overall costs: 
\begin{equation*}
    Minimize \,\,\,\,\, f_2 ( \, \vec{x})\, = \sum_{i=1}^{n} cost_i \cdot x_i  
\end{equation*}

Within the decision vector $\vec{x} = \{x_1,...,x_n\}$, $x_i$ is $1$ if the feature $i$ is selected for the next release and $0$ otherwise.

\subsection{Feature Specification}
\label{sect:featureSpecification}
To support the estimation of concrete values for $value(f_i, s_j)$, we enter the first step in the process outlined in Fig.~\,\ref{fig:steps}. In this step, we collect stakeholder information and convert the expectations into a comprehensive and structured feature specification using the Gherkin 
syntax \cite{SmartBearSoftware}. 
One example is shown in Listing~\ref{list:bspGherkin}. In this way, we can create multiple feature files, where each feature file contains one or more usage scenarios that describe the expected system behavior from a user's point of view. 
\begin{lstlisting}[caption=Feature specification with the help of usage scenarios.,
	label=list:bspGherkin,
	style=GherkinStyle
	]
Feature: User-managed charging (UMC): The user of an electric vehicle requests up-to-date information on energy prices and enters preferences into a smartphone app to calculate an optimized charging plan. 
  Scenario: The EVU requests information on energy prices 
    When the EVU request information on energy prices via the smartphone app
    Then the smartphone app requests these information from an energy information service 
    And the energy information service sends this information to the smartphone app
    And the smartphone app displays the received information 
  Scenario: The EVU user enters charging preferences 
    When the EVU user enters charging preferences
    Then the smartphone app calculates an optimized charging plan
    And the smartphone app sends the charging plan to the electric vehicle
    And the electric vehicle executes this charging plan 
\end{lstlisting}

In the SoS example shown in Fig.~\ref{fig:systemOfSystems}, this step is done by the system owner O1, who directs the operation of the integrated SoS functionality. On this level of abstraction, the individual CSs are seen as black boxes and the usage scenarios are used to document how the CSs must interact to realize the expected SoS functionality. 

By using this kind of feature specification, we can create separate files for each feature $f_i(1 \leq i \leq n)$ to provide an initial overview of the desired system functionality.

\subsection{Definition of Systems and their Interaction}
\label{sect:behaviorModeling}
Based on the initial set of features, we create \emph{scenario specifications} that include definitions of all CSs and how these CSs interact (see. example in Listing \ref{list:bspSMLK}). 
By executing and testing these scenario specifications we successive refine both, the initial feature specifications and the scenario specifications. As outlined in Fig.~\,\ref{fig:valueAndCosts}, this is done by utilizing usage scenarios to link features and technical requirements. 
To bridge the gap from the feature specification to the technical requirements, we generate test steps from the usage scenarios and apply the test-driven scenario specification approach (TDSS) as introduced in previous work \cite{Wiecher2019, Wiecher2020}. 

As outlined in Fig.~\,\ref{fig:smlkexample}, we do this on two levels of abstraction within a SoS context.  
\begin{figure}[h]
    \centering
    \includegraphics[width=1.0\linewidth]{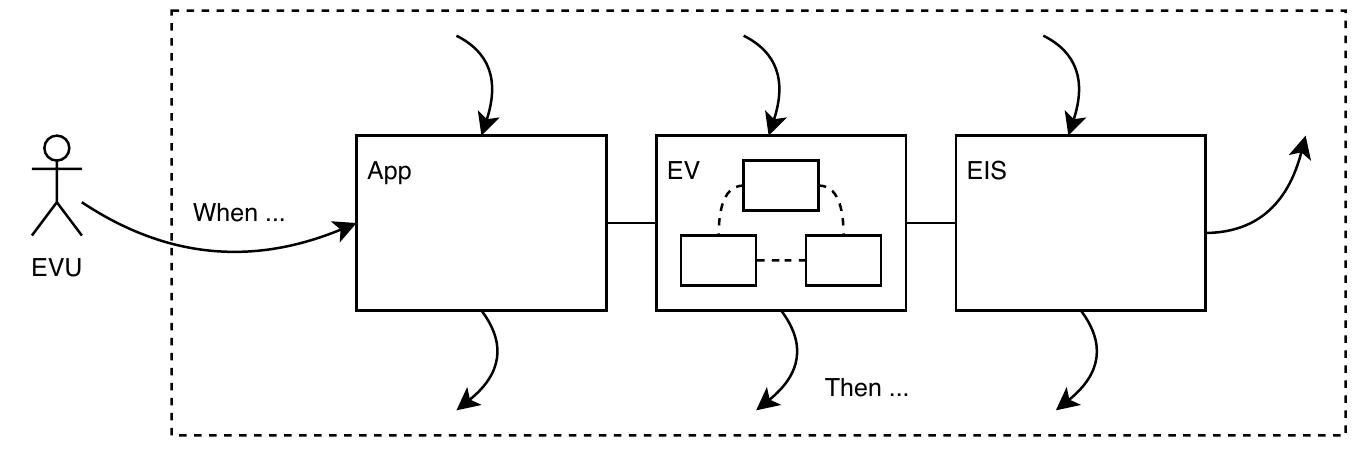}
    \caption{Identified systems with different levels of detail.}
    \label{fig:smlkexample}
\end{figure}
In a first step we define the interaction between CSs, where these systems are seen as black boxes. The focus is to identify the required systems and model their interaction. Subsequently we create separate scenario specifications to model the internal behavior of selected CSs (see.~\cite{Wiecher2021}). 

As an example, to realize the feature specified in Listing~\,\ref{list:bspGherkin}, we can define three systems (smartphone app, electric vehicle, and an energy information service), where an electric vehicle user is interacting with a smartphone app. 
Now we can assume that the smartphone app and the energy information service are existing systems that should be integrated with an electric vehicle under development.  In this case, it is also necessary to identify the components of the electric vehicle that are involved in the desired SoS functionality, which can be done in a separate scenario specification that details the internal behavior of a CS \cite{Wiecher2021}.

\section{Proof of Concept}
To exemplary apply our approach, we integrated SMLK \cite{smlk} with the Cucumber tooling \cite{cucumber} and the MOEA framework \cite{Hadka}. Subsequently we 
executed the steps shown in Fig.~\,\ref{fig:steps} and thereby created the artifact shown in Fig.~\,\ref{fig:profOfConcept} with the numbers 1-4 indicating the associated process step. 
\vspace{-0.2cm}
\begin{figure}[h]
    \centering
    \includegraphics[width=1.0\linewidth]{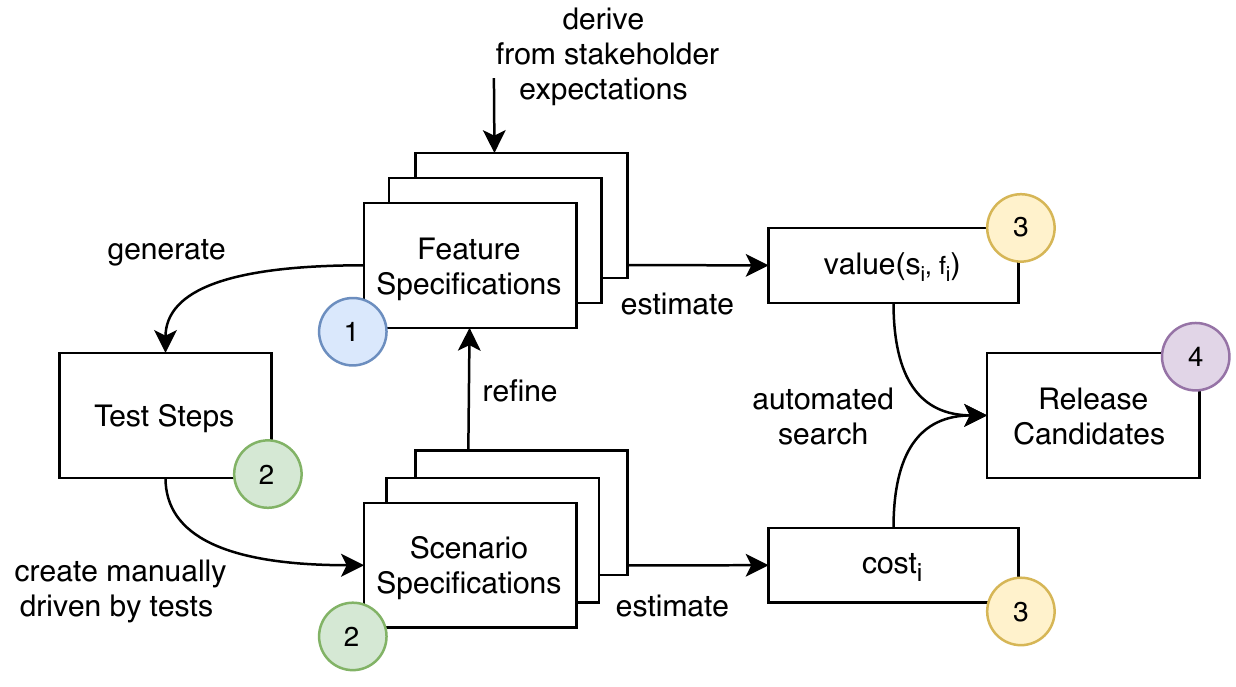}
    \caption{Resulting artifacts when executing the process steps}
    \label{fig:profOfConcept}
\end{figure}

We started with the creation of feature files containing usage scenarios as already shown in Listing \ref{list:bspGherkin}. 
After we created a first set of features, we generated the test steps. 
Based on the example feature in Listing \ref{list:bspGherkin}, we generated the test steps in Listing \ref{list:bspGeneratedTest} and iterated the TDSS sub-process (step 2 in Fig.~\,\ref{fig:steps}). In each iteration we a) added SMLK events to the generated test steps, b) executed the tests, and c) refined the scenario specification until all tests passed. 
\begin{lstlisting}[caption=Generated test step from the usage scenario in Listing \ref{list:bspGherkin}.,
	label=list:bspGeneratedTest,
	style=GherkinStyle
	]
When("^the EVU user enters charging preferences$") {
  trigger(EVU sends App.enterChargingPreferences())} //manually added event 
Then("^ the smartphone app calculates an optimized charging plan$") { 
  receive(App.calculateChargingPlan())} //manually added event 
...
\end{lstlisting}

After several iterations within step 2, the test results could be used to refine the initial feature specifications. Then we used these feature specifications to manually derive the value vector. To do this, we defined hypothetical values based on the amount of usage scenarios for a specific feature and the respective stakeholder. 
In the same way, we derived a cost vector depending on the complexity of the scenario specification with its CSs and their subsystems, representing the technical requirements of a specific SoS.  
\vspace{-0.3cm}
\begin{figure}[h]
    \centering
    \includegraphics[width=1.0\linewidth]{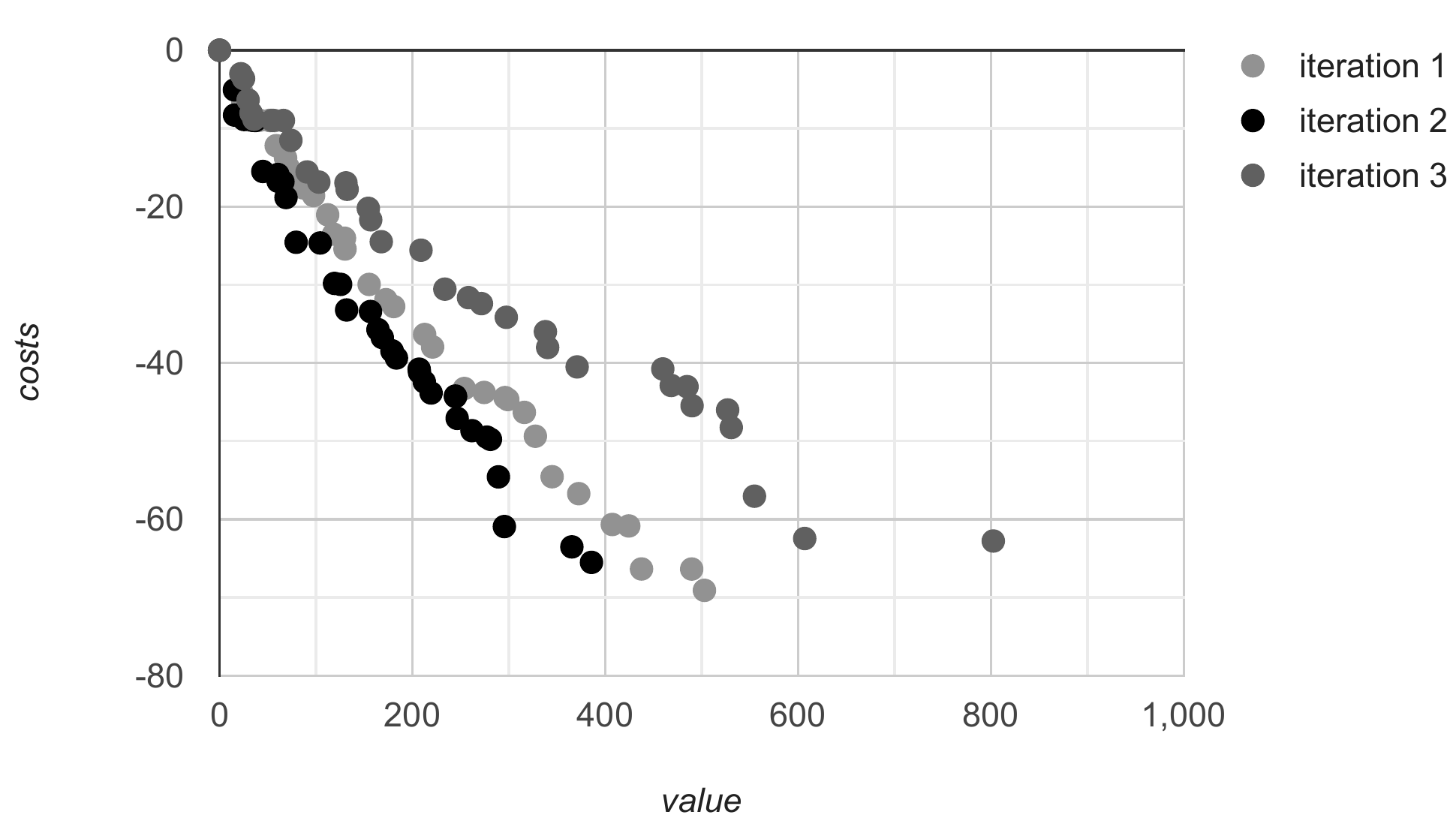}
    \caption{Search results after three iterations with 10 stakeholder and 40 features.}
    \label{fig:esam}
\end{figure}

With this input we exemplary executed the automated search for 10 stakeholder and 40 different features. In Fig.~\,\ref{fig:esam} we see the 
search results with the input data of three iterations. For every iteration we got a pareto optimal front, where each point indicates one possible release candidate. Each release candidate includes a vector containing all features, with 1 if the feature should be selected for the release and 0 otherwise.    

\section{Summary and Outlook}
In this work, we combine a structured feature specification with a scenario-based modeling technique to assist the release engineer in identifying costs and values for the specified features in an early planning phase of an SoS. The cost and value are forming the link into the decision-making processes of the stakeholders in a Smart City since they can be linked to the inputs and the impact of the respective project. 

Feature specifications in natural language are used to support the identification of values that these features can create. Using natural language supports the involvement of the stakeholders into the decision-making. 
The formal and scenario-based modeling of technical requirements is used to support the estimation of costs, depending on the effort it can take to realize the requirements via different CSs. 
We connect both artifacts (feature specification, scenario specification) with a test-driven specification approach based on stakeholder expectations formulated as usage scenarios.   

With this approach we address SoS characteristics like evolving development of individual CSs, where requirements can change constantly but nevertheless can be part of a higher level SoS functionality. The used modeling techniques and methods \cite{Wiecher2019, Wiecher2020, Wiecher2020a, Wiecher2021} support the iterative analysis and refinement of the changed requirements. 
The iterative specification process combined with the automated search for release candidates based on the MONRP is a promising approach to support the release engineer in complex trade-off decisions. 

Future work should consider how to extend the modeling capabilities to estimate more realistic input data for the search. The objective functions should also be adapted to allow the search for features in a realistic release situation. Since this first approach is based on the problem statement in \cite{Zhang2007a}, it is assumed that all features are independent from each other, which is an unrealistic assumption. One possible improvement would be to combine the feature specifications with goal models, as done by Aydemir et al. \cite{Aydemir2018}, and in this way also consider dependencies between features and requirements when searching for optimal release candidates.

\bibliographystyle{ieeetr}
\bibliography{sample-base.bib}
\end{document}